\documentclass[pra,preprint,superscriptaddress]{revtex4-2}
\usepackage{times}
\usepackage[T1]{fontenc}
\usepackage[latin9]{inputenc}
\usepackage{color}
\usepackage{hyperref}
\usepackage{float}
\usepackage{amsmath}
\usepackage{amssymb}
\usepackage{graphicx} 
\usepackage{babel}

\begin{document}
\title{\rev{Efficient calculation of reactor noise via Ito-Langevin Process for correlated fluctuations}}
\author{Guy Gabrieli}
\email{guyga@soreq.gov.il}
\affiliation{School of Physics and Astronomy, Tel Aviv University, Tel Aviv 69978, Israel}
\affiliation{Soreq Nuclear Research Center, Yavne 81800, Israel}
\author{Yair Shokef}
\affiliation{School of Mechanical Engineering, Tel Aviv University, Tel Aviv 69978, Israel}
\affiliation{School of Physics and Astronomy, Tel Aviv University, Tel Aviv 69978, Israel}
\affiliation{Center for Physics and Chemistry of Living Systems, Tel Aviv University, Tel Aviv 69978, Israel}
\affiliation{Center for Computational Molecular and Materials Science, Tel Aviv University, Tel Aviv 69978, Israel}
\author{Izhar Neder}
\affiliation{Soreq Nuclear Research Center, Yavne 81800, Israel}
\def\rev{\textcolor{black}}
\begin{abstract}
We derive an Ito-Langevin stochastic process that captures the time-dependent deviation from Poisson behavior of the noise detected from a general heterogeneous sub-critical neutron system. Using the probability generating function for the actual physical process, we deduce the super-Poisson deviation of the covariance matrix of counts at the detector due to neutron multiplication upon fission. This leads to a general form that coincides with the second moment of an Ito process. This comparison facilitates the formulation of a corresponding effective Langevin equation, which potentially enables simulations that significantly reduce the computational resources required compared to direct simulation of the system's actual noise. This method could assist in designing sub-critical noise experiments for licensing new research reactors, for improving cross-section libraries and for non-destructive assays of spent fuel.
\end{abstract}
\maketitle

\newpage

\section{Introduction} \label{sec:intro}

A prevalent approach to the analysis of counting processes of neutrons in physical systems containing fissile material focuses on deviations of such processes from Poisson statistics. Neutron transport entails a branching process, wherein a single neutron may induce fission, introducing multiple neutrons back into the system. Thus, pairs of counted neutrons that trace back to a common fission event are temporally correlated, giving rise to a super-Poisson counting process. This non-trivial statistics of the counts contains information on the multiplying system, which is often hard to extract from the mean values of the detection rates. Therefore, measuring the statistics of the counts became a valuable tool in the study of nuclear reactors, sub-critical assemblies, accelerator driven systems, nuclear Safeguards applications and more~\cite{diven1956, uhrig1975random,MIHALCZO19771, kim1996design, hergenreder2007kinetic, CROFT2012152, doyle2019nuclear, LEVI2020163797}.

Feynman and de-Hoffman derived a closed expression for the noise behavior of a point-model subcritical system~\cite{feynman1944intensity, feynman1956dispersion}, in which the entire spatial and orientational dependencies are lumped, and the system is described only in terms of the neutron number as function of time. Since then, noise measurements became a standard tool in the commissioning of research reactors~\cite{kim1996design, hergenreder2007kinetic}. However, such reactors are usually highly heterogeneous, and can not always be described by a point model. While in most experiments the Feynman-alpha and Rossi-alpha formulas give a reasonable approximation~\cite{feynman1944intensity, feynman1956dispersion, hergenreder2007kinetic, uhrig1975random,tonoike2004real, kuramoto2005rossi, kuramoto2006rossi, ROSSI_PU_2020}, exceptions exist~\cite{BERGLOF2011194, MUNOZCOBO2011590, SZIEBERTH2015146}. The literature contains several generalized formulas for heterogeneous cases based on the derivatives of the Probability Generating Function (PGF), usually in a form of a sum over two or more eigenmodes of the (adjoint) neutron transport operator~\cite{sheff1966space2,DEGWEKER2016433, Anderson2012, chernikova2013derivation,Two-Exponential_2020,MALINOVITCH2015297}. Practically, however, calculating the full characteristics of noise in a general reactor remains a challenge. This difficulty arises because traditional methods only calculate the fundamental mode of the neutron transport operator, and higher modes are much harder to quantify~\cite{KOPHAZI2012167}. Additionally, using \rev{analogue} Monte Carlo methods~\cite{werner2017mcnp, leppanen2015serpent, romano2015openmc} to directly track stochastic neutron trajectories for calculating fluctuation in the flux at detectors requires excessive computing power. \rev{ Some solutions were conceived for this, either by combining semi-analytical results \cite{NonAnalogousdiff2013} or by using non-analogous algorithms and variance reduction methods \cite{NonAnalogousMC2014}, however their applicability to full-scale reactor problems has not been demonstrated.}

Recently, we demonstrated how source calculations could be utilized in standard Monte Carlo simulations to precisely determine one aspect of detection noise -- its asymptotic behavior over extended time windows~\cite{GABRIELI2023}. Importantly, we identified significant discrepancies between point models and exact position- and energy-dependent calculations for a core configuration of the Israeli Research Reactor 1 (IRR1) in Soreq Nuclear Research Center. This raises the question whether one can calculate the full time-dependent second moment correlations using available core simulation tools, \rev{ which only calculate the first moment, i.e the average (time dependent) flux distribution.}

Here, we develop a simplified method that accurately captures the exact second moments of noise in a heterogeneous system. Specifically, we focus on the correlation between detection events with arbitrary temporal delay between them and possibly at two different locations. Our approach employs stochastic dynamics through a Langevin equation for a Gaussian process, thus leading to a partial differential equation for fluctuations in the neutron flux with a stochastic noise as source. Notably, past efforts have successfully incorporated Ito-Langevin processes in reactor noise modeling\rev{, applying a Schottky-like formula with a noise-equivalent source term}, usually for point models~\cite{cohn1960simplified, Degweker2011296, DUBI2018608}, but also for space-dependent effects~\cite{akcasu1966application, sheff1966space1, sheff1966space2, saito1967noise}. Our approach is substantially different; while in previous works, all stochastic processes in the reactor were accounted for as a source of the random noise, including neutron scatterings, neutron captures, and the detection process itself, we simplify the equation for the full 3D heterogeneous case by incorporating only the branching events -- namely fission -- as source terms. Thus, we formulate a Langevin equation, where the fluctuating source term accounts only for the \textit{deviations} from Poisson statistics. We show that, since other processes generate Poisson noise, the non-Poisson term depends only on the average spatial fission rate distribution, and the distribution of the number of neutrons emitted in a fission event. The resulting equation allows expressing fluctuations using the spatial dependence of average quantities \rev{(i.e. quantities that depend on the first moment of the time-dependent flux)}, which are much simpler to numerically simulate. Thus, our method facilitates the use of simulations to calculate spatial effects of fluctuations. We suggest realizing this fluctuating source using standard Monte Carlo techniques as a simple and accurate way of simulating temporal correlations in the neutron noise. \rev{It should be stated that the treatment we present here does not take into account delayed neutrons: although these can in principle have an effect on noise measurements in neutron systems, we focus on timescales much smaller than the half-lives of delayed neutron precursors, and therefore analyze only prompt neutron effects (in the case of nuclear reactors not too close to criticality, the relevant prompt timescales are of several milliseconds, and  the effects of delayed neutrons become significant much later).}

The outline of the paper is as follows: In Section~\ref{definitions} we give a mathematical description of the quantities of interest - the super-Poisson part of the system's noise and the related Ito-Langevin process. In Section~\ref{Gaussian Eto process}, we express the second moments of the Ito-Langevin process using the one-particle Green function of the transport operator. In Section~\ref{dij} we derive similar expressions for the auto-correlation and cross-correlation of reactor noise, using derivatives of the PGF, and assuming a Poisson external source. This part relies mainly on the expansion of the PGF in Bell's work~\cite{Bell1965} and the resulting expressions. In Section~\ref{comperison}, we compare the expressions found in the two former sections, which enables us to identify the correct Ito-Langevin process for our purpose. In Sec.~\ref{sec:test-case} we solve the Langevin equation for the point-model, as a test case, and show that it leads to the known result \cite{feynman1944intensity,feynman1956dispersion}. Finally, in Section~\ref{discussion} we give some insights on the form of the random source term and on its potential to speed-up computation time of simulations of reactor noise.

\section{Neutron branching process and Langevin equation for its non-Poisson part \label{definitions}}

We start with the variance-to-mean ratio of the number $N_D(S,T)$ of neutrons detected at detector $D$ during the time interval $[0,T]$ near a heterogeneous sub-critical system due to an external stationary Poisson source $S$~\cite{feynman1944intensity, feynman1956dispersion},
\begin{equation}
    \frac{\textrm{Var}(N_D)}{\overline{N_D}}=1+Y(D,T)\label{eq:Fenymann Y} ,
\end{equation}
where $Y$ quantifies the deviation from Poisson statistics at the detector. We generalize this by considering two detectors $D_1$ and $D_2$, and write the following $2\times2$ matrix $d_{ij}$, which represents the deviation from Poisson behavior of the auto-correlation and of the cross-correlation of the numbers of neutrons at the two detectors, 
\begin{align}
   d_{11}(T) &\equiv \textrm{Var}(N_{D_1}) - \overline{N_{D_1}}  \nonumber \\
   d_{22}(T) &\equiv \textrm{Var}(N_{D_2}) - \overline{N_{D_2}} \nonumber \\
   d_{12}(T) &= d_{21}(T) \equiv \overline{N_{D_1}N_{D_2}}-\overline{N_{D_1}}\cdot\overline{N_{D_2}}, 
    \label{eq:d_defined}    
\end{align}
of which Eq.~(\ref{eq:Fenymann Y}) is a special case, as $d_{ii}\rev{(T)}= \overline{N_{D_i}(S,T)}Y(D_i,T)$ for $i=1,2$. The matrix $d_{ij}\rev{(T)}$ is the property of interest in this work. 

The definition of $d_{ij}$ may be generalized by including two different time intervals $T_1$ and $T_2$ for the two detectors. The significance of the general formula is that its mixed derivative with respect to $T_1$ and $T_2$ is the probability \rev{of counting two neutrons that are correlated, namely belong to the same fission chain,} one neutron in detector $D_i$ in the time interval $dt_1$ around $t_1$, and another in detector $D_j$ in the interval $dt_2$ around $t_2$ \rev{(per double time intervals). We denote this function $P_{ij}\left(t_1,t_2\right)$. It} is related to the \rev{correlated part of the} Rossi-alpha function, which is the conditional probability that a neutron is captured within time $\tau$ given an initial count at some time $t$: 
\begin{equation}
    R_{ij}(\tau) =\rev{P_D + }\frac{P_{ij}(t,t+\tau)}{P_D}, \label{eq:Rossi_a_initial}
\end{equation}
where $P_D\rev{d\tau}$ is the probability of a neutron to be detected \rev{at time interval $d\tau$}. Throughout the analysis below we stick with the simpler form in Eq.~(\ref{eq:d_defined}), but show that analyzing such generalizations would be rather straightforward, and we provide the proper formulas for both the Feynman-alpha and the Rossi-alpha methods. \rev{From this point, when $d_{ij}$ is referenced, it should be interpreted as $d_{ij}(T)$.}

Note that the detection noise is by no means Gaussian -- it is in fact almost a Poisson process, and $N$ is strictly non-negative. Nonetheless, we argue that the second moment of the deviation from Poisson behavior, which is captured by $d_{ij}$, has the same behavior as in a Gaussian time-independent Ito-Langevin stochastic process, and that this is true for any choice of the two detectors $D_1$ and $D_2$ and of the time interval $T$. 

We denote the single-neutron phase space as $\vartheta\equiv\left(\mathbf{x},E,\mathbf{\Omega}\right)$, where $\mathbf{x}$ is the three-dimensional neutron position in the reactor, $E$ is the neutron energy and $\mathbf{\Omega}$ is the neutron direction of motion. We consider a stochastic function $\delta n\left(\vartheta,t\right)$ defined on this phase space, that represents -- up to the second moment -- the part of the fluctuation of the neutron number density that deviates from Poisson behavior. Our Ito-Langevin process is defined by \rev{the following ansatz,}
\begin{equation}
\hat{L}_\vartheta\delta n\left(\vartheta,t\right)=\sqrt{s(\mathbf{x})}\chi\left(\mathbf{\vartheta}\right)\epsilon(\mathbf{x},t)\label{eq:Langevin},
\end{equation}
where $\hat{L}_\vartheta$ is a transport operator that depends on $\vartheta$ (the full transport operator is defined below, but the method applies for any substitute, such as the multi-group diffusion operator), $s(\mathbf{x})$ and $\chi(\mathbf{\vartheta})$ are time independent functions, that determine the noise fluctuation strength and spectrum (in both energy and direction in each spatial point), respectively, and $\epsilon(\mathbf{x},t)$ is uncorrelated Gaussian white noise with zero mean, and with variance unity,
\begin{equation}
 \overline{\epsilon(\mathbf{x},t)\epsilon(\mathbf{x'},t)} = \delta \left(t-t'\right) \delta^3 \left(\mathbf{x}'-\mathbf{x}\right) ,
 \label{eq:epsilon_def}
\end{equation}
where $\delta$ is the Dirac delta function. Since the transport operator $\hat{L}_\vartheta$ is typically linear, the solution $\delta n$ will itself be a Gaussian process, as it is given by a convolution of a Gaussian white noise.

Our aim is to find a Langevin equation of the form of Eq.~(\ref{eq:Langevin}), such that its solution will reproduce the exact second moment \rev{of the noise. Namely, we look for the noise amplitude $s(\mathbf{x})$ and spectrum $\chi(\mathbf{\vartheta})$ that will correctly recreate the second moment matrix $d_{ij}(T)$ defined in Eq.~(\ref{eq:d_defined}).} To this end, we define the detector~$D$ through a detection cross section $\sigma_D(\vartheta,t)$ which is non-zero only in a localized phase-space region around the detector's position, orientation and neutron energy sensitivity and is non-zero only in the time interval $[0,T]$. Upon detection, we assume that $\sigma_D(\vartheta)$ necessarily causes a capture of a neutron, although there are detectors that use neutron scattering events or fission chambers, in which neutrons emerge from the detector after it ``clicked''~\cite{cknoll2010radiation}. Such detectors produce additional non-Poisson auto-correlations that do not stem from the fission chains in the core, which are the source of the noise that we are interested in. To avoid this complication, these kind of detectors are excluded in our theoretical discussion. 

Given the effective Gaussian noise $\delta n$ obeying Eq.~(\ref{eq:Langevin}), the detection signal fluctuation that it causes at detector $D$ at time $t\in[0,T]$ is given by 
\begin{equation}
    \delta n_{D}(t) \equiv \int_D d\vartheta \sigma_D(\vartheta,t) v(E) \delta n(\vartheta,t) , \label{eq:delta n_D}
\end{equation}
where $v(E)$ is the neutron velocity and $d\vartheta=d\mathbf{x}dEd\mathbf{\Omega}$ is the phase space differential.  

The key point of our work is to compare the covariance matrix of these effective signals with a derivation of $d_{ij}(T)$ for a general reactor geometry and to show that
\begin{equation}
   d_{ij}(T)=\int_0^T dt_1 \int_0^T dt_2\overline{\delta n_{D_{1}}(t_{1})\delta n_{D_{2}}(t_{2})} , \label{eq:aim_of_paper}
\end{equation}
provided that the fluctuating source of the Langevin equation~(\ref{eq:Langevin}) is characterized by
\begin{equation}
    s(\mathbf{x})=\overline{\nu(\nu-1)}\cdot\overline{F(\mathbf{x})} , \label{eq:source}
\end{equation}
where $F(\mathbf{x})$ is the average fission rate at steady state, and $\nu$ is the random variable for the number of neutrons emerging from a fission event, and that $\chi\left(\mathbf{\vartheta}\right)$ in Eq.~(\ref{eq:Langevin}) is the fission neutrons' spectrum (this formulation holds for any nuclear reaction producing multiple neutrons. In the case of fission, which can be considered isotropic, we have $\chi\left(\mathbf{\vartheta}\right)=\frac{\chi\left(E\right)}{4\pi}$). That is to say, a simple Langevin equation, with a Gaussian white noise resulting from the fission rate distribution, recreates the correct deviation from Poisson behavior of the auto- and cross-correlations of the neutron counting process at the detectors.
 
\begin{figure}[b]
\centering 
\begin{tabular}{|c|}
\hline 
\includegraphics[scale=0.4]{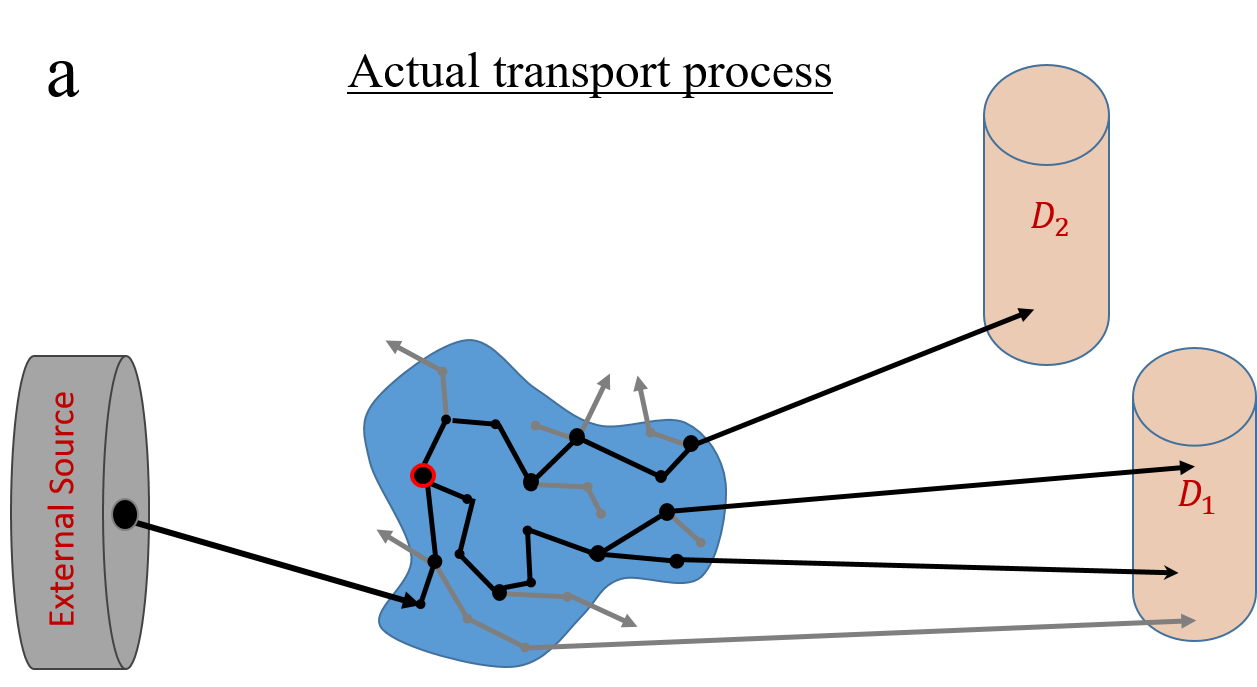}\tabularnewline
\hline 
\hline 
\includegraphics[scale=0.4]{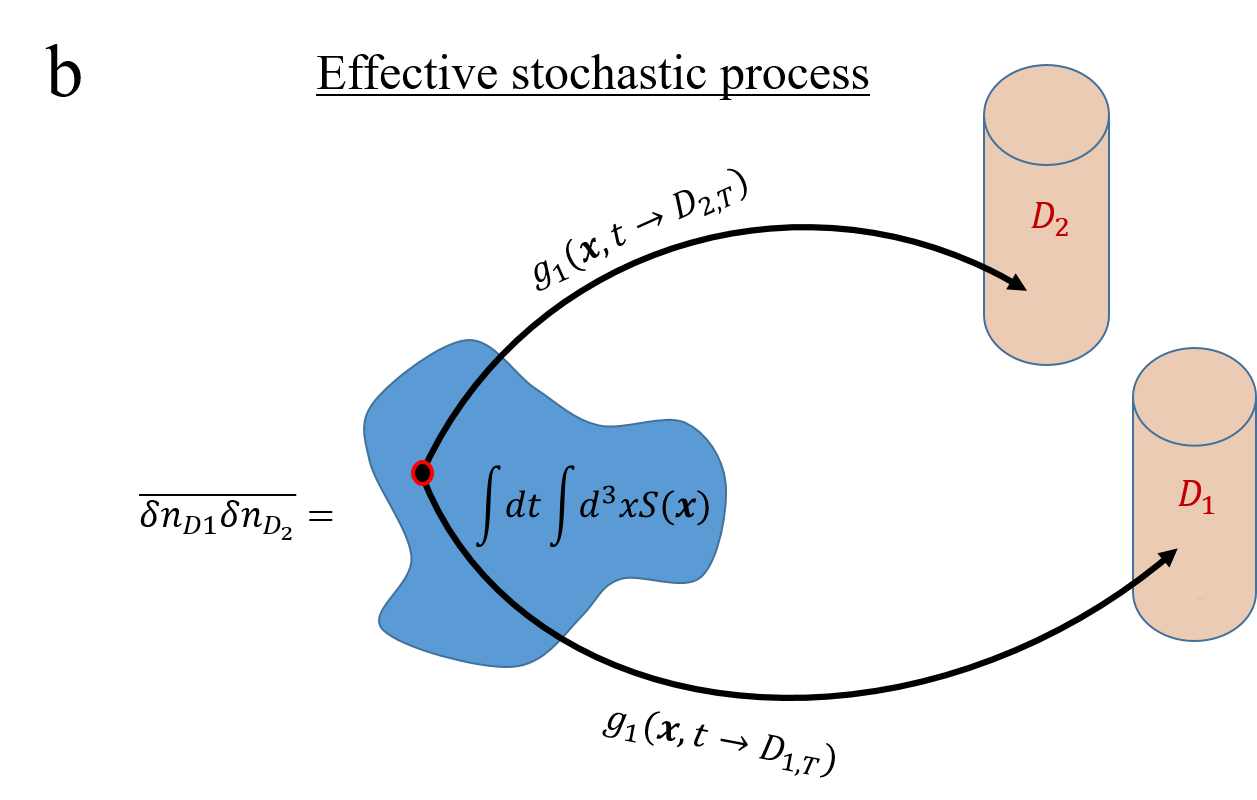}\tabularnewline
\hline 
\end{tabular}
\caption{Equivalence between non-Poisson correlations of the subcritical system detection signal and those originating from a Langevin process. (a) Deviation from Poisson statistics in the detectors is due to pairs of neutrons from the same fission chain. Pairs originate from all possible neutron histories starting in a particular external source event, and can be counted by assigning them to a distinct fission event (red circle) from which their trajectories branched, eventually reaching the detectors (black line). Other detected neutrons (gray lines) will be paired and counted by going over all fission events. (b) Feynman-diagram showing how the cross-correlation of a Langevin process can be expressed as an integral over the noise source within the system (red circle) and two Green functions (black lines) to the detectors.}
\label{fig:Feynmann_Langevin}
\end{figure}

Figure~\ref{fig:Feynmann_Langevin} demonstrates the reasoning behind the \rev{ identity in Eq.~(\ref{eq:aim_of_paper}) between the second moment of the noise of the full stochastic transport picture and that produced in the effective Langevin treatment we present}. The deviation $d_{ij}\left(T\right)$ from Poisson behavior can be attributed to pairs of detected neutrons which originated from the same fission chain~\cite{feynman1944intensity}. The key point is a method of counting all those pairs, by first assigning each pair to their last common fission event, and then counting the pairs by going over all fission events. This is depicted in Fig.~\ref{fig:Feynmann_Langevin}a: for each fission event (one of which is marked in red) we count all the neutrons detected in the detectors $D_1$ and $D_2$, whose histories (the black lines) branched from this fission event. Counting all such pairs and summing over all fission events leads to $d_{ij}$. With this order of summation, both sides of Eq.~(\ref{eq:aim_of_paper}) can be expressed by a Feynman-diagram-like expression, which is depicted in Fig.~\ref{fig:Feynmann_Langevin}b: it involves a product of two Green functions $g_1$, which describe the transport of neutrons originating from the fission event at $\mathbf{x}$ to the two detectors, and then an integration over all such fission events. For the calculation of $d_{ij}$ it is a result of the above counting method of pairs of neutrons from the same fission chain, whereas for the Ito-Langevin process it is a result of its Gaussian characteristic. Below we derive, for each side of Eq.~(\ref{eq:aim_of_paper}) separately, an expression of such a double-Green-function form. The comparison between the two resulting expression leads to the exact form of Eqs.~(\ref{eq:Langevin}, \ref{eq:source}). Importantly, this directly leads to a more efficient way to calculate the noise correlation via simulations.

\section{Gaussian Ito process transported in time and space\label{Gaussian Eto process}}

The fact that one can express the second moment of the above Ito-Langevin process using propagators should not come as a surprise, as this is a general property of any stationary Gaussian process. Here we derive the result for our process of interest; we have a steady state of a stationary stochastic system, whose state is described by a random field with zero mean $\delta n(\vartheta,t)$ and which obeys Eq.~(\ref{eq:Langevin}). To be concrete, we take $\hat{L_\theta}$ to be the neutron transport operator, 
\begin{equation}
    \hat{L_\vartheta}\equiv\frac{\partial}{\partial t}+\mathbf{v}\cdot\mathbf{\nabla}+v\sigma(\vartheta)-v\sigma(\vartheta)\int d\mathbf{\Omega}'dE'\overline{c}\left(\vartheta'_\mathbf{x}\rightarrow\vartheta_\mathbf{x}\right), \label{eq:L_def}
\end{equation}
where $\sigma\left(\vartheta\right)$ is the total cross section for capture, scattering and fission, and $\overline{c}\left(\vartheta'_\mathbf{x}\rightarrow\vartheta_\mathbf{x}\right)$
denotes the average number of neutrons, transitioning due to collision and fission, from phase-space point $\vartheta'_\mathbf{x}=\left(\mathbf{x},\mathbf{\Omega'},E'\right)$ into point $\vartheta_\mathbf{x}=\left(\mathbf{x},\mathbf{\Omega },E\right)$. The subscript $\mathbf{x}$ in $\vartheta_\mathbf{x}$ and $\vartheta'_\mathbf{x}$ indicates that both are at the same position $\mathbf{x}$, but with different energies and orientations. \rev{Throughout this work, integration is performed over the full range permissible for the variable, unless the integration range is explicitly stated.}
In the next section we also use its adjoint operator,
\begin{equation}
    \hat{L^\dagger_\vartheta}\equiv-\frac{\partial}{\partial t}-\mathbf{v}\cdot\mathbf{\nabla}+v\sigma(\vartheta)-v\sigma(\vartheta)\int d\mathbf{\Omega}'dE'\overline{c}\left(\vartheta_\mathbf{x}\rightarrow\vartheta'_\mathbf{x}\right). \label{eq:L_dag_def}
\end{equation}
The retarded one-particle Green function $g_1\left(\vartheta',t'\to\vartheta,t\right)$ is formally defined as the inverse of $\hat{L^\dagger}$ and therefore obeys,
\begin{equation}
 \hat{L}_{\vartheta'}^\dagger g_{1}\left(\vartheta',t'\to\vartheta,t\right)=\hat{L}_{\vartheta} g_{1}\left(\vartheta',t'\to\vartheta,t\right)=\delta\left(t-t'\right)\delta\left(\vartheta-\vartheta\right) , \label{eq:Lg1}
\end{equation}
with $g_{1}=0$ for $t'<t$. The formal solution for $\delta n$ in Eq.~(\ref{eq:Langevin}) can then be expressed by a convolution of the Green function $g_1$ with the stochastic source,
\begin{equation}
\delta n(\vartheta,t)=\int d\vartheta'\int_{-\infty}^{t} dt'\sqrt{s\left(\mathbf{x'}\right)}\chi\left(\mathbf{\vartheta}'\right)\epsilon(\mathbf{x}',t')g_{1}\left(\vartheta',t'\rightarrow\vartheta,t\right).   \label{eq:delta_n} 
\end{equation}
We wish to calculate the steady-state correlation between fluctuations in two different volumes in phase space; one at phase-space volume $D_1$ and the other at $D_2$.  We insert Eq.~(\ref{eq:delta_n}) in Eq.~(\ref{eq:delta n_D}), and use the correlations in Eq.~(\ref{eq:epsilon_def}), to arrive, after some algebra to
\begin{align}
\overline{\delta n_{D_{1}}(t_{1})\delta n_{D_{2}}(t_{2})}=& \int_{D_{1}}\sigma_{D_1}v_1d\vartheta_1\int_{D_{2}} \sigma_{D_2}v_2d\vartheta_2\int d\mathbf{x }dEdE'd\mathbf{\Omega }d\mathbf{\Omega }'dt s\left(\mathbf{x}\right)\chi\left(\mathbf{x},E,\mathbf{\Omega }\right)\chi\left(\mathbf{x},E',\mathbf{\Omega }'\right)\cdot\nonumber\\
 & \cdot g_{1}\left(\mathbf{x },E,\mathbf{\Omega },t\rightarrow\vartheta_1,t_{1}\right)g_{1}\left(\mathbf{x },E',\mathbf{\Omega }',t\rightarrow\vartheta_2,t_{2}\right)\label{eq:langevin form}
\end{align}

This expression holds both for the cross-correlation and for the auto-correlation, which is the special case $D_{1}=D_{2}$. Note that the expression in Eq.~(\ref{eq:langevin form}) is in the form depicted in the diagram in Fig.~\ref{fig:Feynmann_Langevin}b, where integration is performed on the source region and the two detector regions. Moreover, we can define the one-particle Green function to any of the detectors $D_i$, $i=1,2$. \rev{Keeping the notation $g_1$ for the any one-particle Green function, we change only the notation of the inner variables, and define it as} 
\begin{equation}
    g_1(\vartheta,t\to D_{i},t')\equiv  \int d\vartheta'\sigma_{D_i}(\vartheta')v' g_{1}\left(\vartheta,t\rightarrow\vartheta',t'\right).\label{eq:g_1_Dt}
\end{equation}
The integral of this Green function over the time period $[0,T]$ \rev{is defined as another Green function,}
\begin{equation}
    g_1(\vartheta,t\to D_{i,T})\equiv \int_0^T dt' g_1(\vartheta,t\to D_{i},t').\label{eq:g_1_D}
\end{equation}
In addition, we shall use the notation $\vartheta'_{1\mathbf{x}}$, $\vartheta'_{2\mathbf{x}}$ and $\int d\vartheta'_{1\mathbf{x}}d\vartheta'_{2\mathbf{x}}$ to indicate integration over energy and direction at a fixed position. Using these definitions and notations, Eq.~(\ref{eq:langevin form}) becomes
\begin{align}
&\rev{\int_0^T}\int_0^T dt_1dt_2\overline{\delta n_{D_{1}}(t_{1})\delta n_{D_{2}}(t_{2})}= \nonumber\\
&=\int d\mathbf{x}dt\int d\vartheta_{1\mathbf{x}}d\vartheta_{2\mathbf{x}} s\left(\mathbf{x}\right)\chi\left(\vartheta_{1\mathbf{x}}\right)\chi\left(\vartheta_{2\mathbf{x}}\right) \cdot g_1\left(\vartheta_{1\mathbf{x}},t\to D_{1,T}\right)g_{1}\left(\vartheta_{2\mathbf{x}},t\to D_{2,T}\right) , \label{eq:langevin form2}
\end{align}
which is the expression for the r.h.s.\ of Eq.~(\ref{eq:aim_of_paper}). 

Note that the two time integrations in Eq.~(\ref{eq:g_1_D}) are the last step of this derivation and therefore \rev{do not} change the form of the results. Therefore, this can be generalized to different time periods in $\sigma_{D_1}(\vartheta,t)$ and $\sigma_{D_2}(\vartheta,t)$. As mentioned in Section~\ref{definitions}, the probability \rev{density} function of \rev{correlated neutron pair} detection can be derived from the result above, by changing $\iint_0^T dt_1dt_2\overline{\delta n_{D_{1}}(t_{1})\delta n_{D_{2}}(t_{2})}$  in Eq.~(\ref{eq:langevin form2}) to $\int_{0}^{T_1}\int_{0}^{T_2} dt'dt''\overline{\delta n_{D_{1}}(t')\delta n_{D_{2}}(t'')}$, differentiating with respect to $T_1$ and $T_2$ and evaluating the result at times $t_1, t_2$. This results in:
\begin{align}
    P_{ij}(t_1,t_2) = 
    & \int_{-\infty}^{\infty} dtd\mathbf{x}\int d\vartheta_{1\mathbf{x}}d\vartheta_{2\mathbf{x}} s\left(\mathbf{x}\right)\chi\left(\vartheta_{1\mathbf{x}}\right)\chi\left(\vartheta_{2\mathbf{x}}\right) \cdot \nonumber \\ &\cdot  g_1(\vartheta_{1\mathbf{x}},t\to D_{i},t_1) g_1(\vartheta_{2\mathbf{x}},t\to D_{j},t_2). \label{eq:Joint_prob}
\end{align}

\section{Detection noise in a general sub-critical assembly \label{dij}}

For the l.h.s. of Eq.~(\ref{eq:aim_of_paper}), we follow the path of many works before us, dating back to~\cite{Bell1965}, and use the PGF to calculate the second factorial moment of the reactor noise at the detectors, in the presence of a stable external Poisson source with average emission rate density $S(\vartheta)$ neutrons per unit volume per unit time, at phase space point $\vartheta\equiv(\mathbf{x},\mathbf{\Omega}, E)$. We are looking for the PGF of the number of neutron detections at our two detectors $D_1$ and $D_2$ during the time interval $[0,T]$. Due to the Poisson nature of the external source, it is a PGF of a compound Poisson process, that is given by~\cite{Bell1965}
\begin{equation}
G(z_{1},z_{2},S)=\exp\left[\int d\mathbf{x}d\mathbf{v}dtS(\vartheta,t)\left(G_{\rm1n}\left(z_{1},z_{2},\vartheta,t\right)-1\right)\right] , \label{eq:compound possion}
\end{equation}
where $G_{\rm1n}(z_{1},z_{2},\vartheta,t)$ is defined as the joint PGF of the probability distribution $P(n_1,n_2|\vartheta,t)$ that a single initial neutron at phase space point $\vartheta$ at time $t$ would cause $n_1$ and $n_2$ neutron detections at $D_1$ and $D_2$, respectively, during the time window $[0,T]$:
\begin{equation}
G_{\rm1n}(z_{1},z_{2},\vartheta,t)\equiv\sum_{n_1=0}^\infty\sum_{n_2=0}^\infty P(n_1,n_2|\vartheta,t)z_1^{n_1}z_2^{n_2}.
\end{equation}
By this definition, and with the normalization of probabilities, we have
\begin{equation}
   \left.G(z_{1},z_{2},S)\right|_{z_{1}=z_{2}=1}= \left.G_{\rm1n}(z_{1},z_{2},\vartheta, t)\right|_{z_{1}=z_{2}=1}=1 .
\end{equation}
The first derivatives \rev{with respect to $z_i$} give the means of the counts at the detectors $i=1,2$ during the time interval,
\begin{equation}
    \overline{N_{D_{i}}(S,T)}=\left. \frac{\partial}{\partial z_i}G(z_{1},z_{2},S)\right|_{z_{1}=z_{2}=1} ,
\end{equation}
the second derivatives give both the cross-correlation,
\begin{equation} 
C_{D_{1}D_{2}}\left(S\right)\equiv\left.\overline{N_{D_{1}}N_{D_{2}}}=\frac{\partial^{2}}{\partial z_{1}\partial z_{2}}G(z_{1},z_{2},S)\right|_{z_{1}=z_{2}=1} \label{eq:Cross_corelation_S}
\end{equation}
and the second-factorial moments for the auto-correlation,
\begin{equation}
C_{D_{i}D_{i}}\left(S\right)\equiv\overline{N_{D_{i}}\left(N_{D_{i}}-1\right)}=\left.\frac{\partial^{2}}{\partial z_{i}^{2}}G(z_{1},z_{2},S)\right|_{z_{1}=z_{2}=1}. \label{eq:C_D1D1}
\end{equation}

Using these relations we obtain the matrix of deviation from Poisson behavior, defined in Eq.~(\ref{eq:d_defined}),
\begin{equation}
    d_{ij}\equiv C_{D_{i}D_{j}}-\overline{N_{D_{i}}}\cdot\overline{N_{D_{j}}} ,
\end{equation}
and from Eqs.~(\ref{eq:compound possion}-\ref{eq:C_D1D1}), we find,
\begin{equation}
   d_{ij}=\int_0^T dt\int d\vartheta S(\vartheta,t)\left.\frac{\partial^{2}}{\partial z_{i}\partial z_{j}}G_{\rm1n}(z_{1},z_{2},\vartheta,t)\right|_{z_{1}=z_{2}=1} . \label{eq:dij}
\end{equation}

Next, we use this relation to express $d_{ij}$ using the one-particle Green function $g_{1}(\vartheta,t\to D_{i,T})$ defined through Eq.~(\ref{eq:g_1_D}). First, note that $g_1$ is the average contribution of a single source neutron at $\vartheta$ at time $t$ to the neutrons detected at detector $D$ within the time interval $[0,T]$. Therefore, by definition, it is given by the derivative of $G_{\rm1n}$,
\begin{equation}
    g_{1}(\vartheta,t\to D_i,t_f)=\left.\frac{\partial}{\partial z_{i}}G_{\rm1n}(z_{1},z_{2},\vartheta, t_f-t)\right|_{z_{1}=z_{2}=1}.
\end{equation}
Next, we return to the transport equation that  $G_{\rm1n}\left(z_{1},z_{2},\vartheta,t\right)$ satisfies. Following the steps in~\cite{Bell1965}, the equation is given by 
\begin{align}
&\hat{L}_\vartheta^{\dagger}G_{\rm1n}+v\sigma(\vartheta,t)\int d\vartheta'_{\mathbf{x}}\overline{c}\left(\vartheta\rightarrow\vartheta'_{\mathbf{x}}\right)G_{\rm1n}=\nonumber\\  =&v\sigma_{D_1}(\vartheta,t)z_1+v\sigma_{D_2,t}(\vartheta)z_2+v\sigma(\vartheta)
\cdot\left[\tilde{c}_{0}\left(\vartheta,t\right)+\int d\vartheta'_{1\mathbf{x}}c_{1}\left(\vartheta\rightarrow\vartheta_{1\mathbf{x}}\right)G_{\rm1n}\left(z_{1},z_{2},\vartheta_{1\mathbf{x}},t\right) 
+\right.\nonumber\\ &\left. +\int d\vartheta'_{1\mathbf{x}} d\vartheta'_{2\mathbf{x}}c_{2}\left(\vartheta\rightarrow \vartheta'_{1\mathbf{x}},\vartheta'_{2\mathbf{x}}\right)G_{\rm1n}\left(z_{1},z_{2},\vartheta'_{1\mathbf{x}},t\right)G_{\rm1n}\left(z_{1},z_{2},\vartheta'_{2\mathbf{x}},t\right)+...\right],&  \label{eq:Full G1 equation}
\end{align}
where $\hat{L}_\vartheta^\dagger$ is the adjoint transport operator defined in Eq.~(\ref{eq:L_dag_def}) above, $c_i\left(\vartheta\to\vartheta_{1\mathbf{x}},..,\vartheta_{i\mathbf{x}}\right)$ is the fraction of the total cross-section at phase-space point $\vartheta=(\mathbf{x},\mathbf{\Omega},E)$ for creation of $i$ neutrons at points $\vartheta_{1\mathbf{x}},...,\vartheta_{i\mathbf{x}}$, all at position $\mathbf{x}$, and $\tilde{c}_0(\vartheta,t)$ denotes the captures excluding the detection, i.e. $\sigma\tilde{c}_0=\sigma c_0-\sigma_{D_1} -\sigma_{D_2}$. The future boundary condition needed for Eq.~(\ref{eq:Full G1 equation}) to complete the noise problem is $G_{\rm1n}\left(z_1,z_2,\vartheta,t\right)=1$ for $t>T$. The equation can be understood by integration backward in time from this future condition toward the past. As the only explicit $z_1$ and $z_2$ dependence is in the two detecting regions $D_1$ and $D_2$ at time period $[0,T]$, the detectors acts as the sources of this adjoint problem.  

Taking the first derivative with respect to $z_{i}$ at $z_1=z_2=1$ results in the adjoint transport equation of $g_{1}$, similar to Eq.~(\ref{eq:Lg1}) but for a finite detection region, where for $t<T$, 
\begin{equation}
\hat{L}_\vartheta^{\dagger}g_{1}(\vartheta,t\rightarrow D_{i,T})=v\sigma_{D_i}(\vartheta,t) , \label{eq:g1_D_eq} 
\end{equation}
with the final condition
\[
g_{1}\left(\vartheta,t\rightarrow D_{i,T}\right)=0,\;\; t>T .
\]
One can see that the definition in Eq.~(\ref{eq:g_1_D}) indeed leads to  Eq.~(\ref{eq:g1_D_eq}).  
Taking the second derivatives of Eq.~(\ref{eq:Full G1 equation}) above with respect to any combination of $z_{1}$ and $z_{2}$, we find 
\begin{align}
&\hat{L}_\vartheta^{\dagger}\left[\left.\frac{\partial^{2}}{\partial z_{i}\partial z_{j}}G_{\rm1n}(z_{1},z_{2},\vartheta,t)\right|_{z_{1}=z_{2}=1}\right]=\nonumber\\&v(\vartheta_\mathbf{x})\sigma(\vartheta_\mathbf{x})\cdot\sum_{i=2}^{\infty}\int d\vartheta'_{1\mathbf{x}} d\vartheta'_{2\mathbf{x}}c_{i}\left(\vartheta_\mathbf{x}\rightarrow\vartheta'_{1\mathbf{x}},\vartheta'_{2\mathbf{x}}\right)\cdot\nonumber\\
&\cdot i(i-1)g_{1}(\vartheta'_{1\mathbf{x}},t\rightarrow D_{1,T})g_{1}(\vartheta'_{2\mathbf{x}},t\rightarrow D_{2,T})&\label{eq:L_dag_CD1D2} .
\end{align}

The r.h.s. of Eq.~(\ref{eq:L_dag_CD1D2}) already looks like what we need for the Langevin process -- some integral of $g_{1}\cdot g_{1}$. But in order to use it in Eq.~(\ref{eq:dij}) we need to make the l.h.s. of both equations equal. We use the fact that $g_1$ behaves as the inverse of $\hat{L}_\vartheta^{\dagger}$, in the sense that for every function of the phase space, $f(\vartheta,t)$, using integration by parts and Eq.~(\ref{eq:Lg1}), we have,
\begin{equation}
    \int dt d\vartheta g_1(\vartheta',t'\to\vartheta,t) \left[\hat{L}_\vartheta^{\dagger} f(\vartheta,t)\right] =\int dt d\vartheta \left[\hat{L}_\vartheta g_1(\vartheta',t'\to\vartheta,t)\right] f(\vartheta,t)= f(\vartheta',t').
\end{equation} 
We therefore multiply Eq.~(\ref{eq:L_dag_CD1D2}) by $S(\vartheta') g_1(\vartheta',t'\to\vartheta,t)$, and integrate over $\vartheta$, $\vartheta'$ and $t$. The l.h.s. of the result becomes equal to the r.h.s. of Eq.~(\ref{eq:dij}), and so we find,
\begin{align}
&d_{ij}=\rev{\int_{-\infty}^{\infty} dt}\int d\mathbf{x}\int d\vartheta_{1\mathbf{x}} d\vartheta_{2\mathbf{x}}\left[\int d\vartheta_\mathbf{x}\bar{\psi}(\vartheta_\mathbf{x})\sigma(\vartheta_\mathbf{x})\sum_{i=2}^{\infty}c_{i}\left(\vartheta\rightarrow\vartheta_{1\mathbf{x}},\vartheta_{2\mathbf{x}}\right)i(i-1)\right]\cdot\nonumber\\ &\cdot g_{1}(\vartheta_{1\mathbf{x}},t\rightarrow D_{1,T})g_{1}(\vartheta_{2\mathbf{x}},t\rightarrow D_{2,T})  ,  \label{eq:d_ij_g1_g1}
\end{align}
where $\bar{\psi}(\vartheta)$ is the neutron flux at steady state, which is given by
\begin{equation}
\bar{\psi}(\vartheta)=v(\vartheta)\int d\vartheta'\int_{-\infty}^{0} dt' S(\vartheta')g_{1}(\vartheta',t'\to\vartheta,t=0).  
\end{equation}
The r.h.s. of Eq.~(\ref{eq:d_ij_g1_g1}) has a clear physical meaning: it is the average number of possible pairs of neutrons that are detected in $D_1$ and $D_2$ within the time period $[0,T]$, and whose fission chains originate from the same fission event. As each such possible neutron pair has one distinct fission event which is the last event from which the pair's two fission chains originate, the r.h.s. sums over all fission events and counts the average number of detected pairs from two fission chains that originate from this event. With this respect, the factor $i(i-1)$, which is the origin of the factor $\overline{\nu(\nu-1)}$ in Eq.~(\ref{eq:source}) counts the average number of possible pairs of fission chains starting from a particular fission event.

\section{Ito-Langevin equation for the non-Poisson part of the reactor noise\label{comperison}}

Comparing the r.h.s.\ of Eq.~(\ref{eq:d_ij_g1_g1}) and Eq.~(\ref{eq:langevin form2}), they become identical if and only if the source term in Eq.~(\ref{eq:langevin form2}) is the expression in the square brackets in Eq.~(\ref{eq:d_ij_g1_g1}), namely if
\begin{equation}    s\left(\mathbf{x}\right)\chi\left(\vartheta_{1\mathbf{x}}\right)\chi\left(\vartheta_{2\mathbf{x}}\right)=\int d\vartheta_\mathbf{x}\bar{\psi}(\vartheta_\mathbf{x})\sigma(\vartheta_\mathbf{x})\sum_{i=2}^{\infty}c_{i}\left(\vartheta\rightarrow\vartheta_{1\mathbf{x}},\vartheta_{2\mathbf{x}}\right)i(i-1).\label{eq:comparison}
\end{equation}
Consider a spatial distribution of fissile material with time-independent macroscopic fission cross section $\Sigma_f$, and with a distribution $P(\nu)$ of the number of neutrons emitted from fission $\nu$, each emitted with the same spectrum probability distribution $\chi(\vartheta)$ for each outgoing neutron, independently, i.e. the case where 
\begin{equation}
c_{i}\left(\vartheta\rightarrow\vartheta_{1},..,\vartheta_{j} \right)=P(i)\chi(\mathbf{\vartheta}_1)\cdot...\cdot\chi(\mathbf{\vartheta}_j)\frac{i!}{j!(i-j)!}.
\end{equation}
We are interested only in the $j=2$ case, which is the only case appearing in the r.h.s. of Eq.~(\ref{eq:comparison}), and that we can now rewrite as
\begin{equation}    
s\left(\mathbf{x}\right)\chi\left(\vartheta_{1\mathbf{x}}\right)\chi\left(\vartheta_{2\mathbf{x}}\right)=\left[\overline{\nu(\nu-1)}\int d\vartheta_\mathbf{x}\bar{\psi}(\vartheta_\mathbf{x})\sigma_f(\vartheta_\mathbf{x})\right]\chi\left(\vartheta_{1\mathbf{x}}\right)\chi\left(\vartheta_{2\mathbf{x}}\right).
\end{equation}
Note that the term $\int d\vartheta_\mathbf{x}\bar{\psi}(\vartheta_\mathbf{x})\sigma_f(\vartheta_\mathbf{x})\equiv\overline{F(\mathbf{x})}$  is the mean fission rate density at steady state. We can therefore extract from this equation the source term $s(\mathbf{x})$ for the Langevin equation, Eq.~(\ref{eq:Langevin}), which now reads 
\begin{equation}
    \hat{L}_\vartheta\delta n\left(\vartheta,t\right)=\sqrt{\overline{\nu(\nu-1)}\cdot\overline{F(\mathbf{x})}}\;\chi\left(\mathbf{\vartheta}\right)\epsilon(\mathbf{x},t) , \label{eq:Langevin_final}
\end{equation} 
and this proves Eq.~(\ref{eq:source}). 

Equation~(\ref{eq:Langevin_final}) is the main result of this work. Simulating neutron flux fluctuations according to Eq.~(\ref{eq:Langevin_final}) (i.e. fluctuations that originate from a realization of the Gaussian noise source in its r.h.s.) has statistics such that the covariance matrix of the counts at any two detectors that utilize neutron capture for detection matches the \textit{deviation} $d_{ij}$ of the noise of real counts from Poisson behavior.

The probability \rev{density} function of detection \rev{of correlated pairs} can now be derived either directly from Eq.~(\ref{eq:Joint_prob}), or from a generalization of Eq.~(\ref{eq:d_ij_g1_g1}) in a similar manner that led to Eq.~(\ref{eq:Joint_prob}), namely by changing the two $T$'s into $T_1$ and $T_2$, differentiating twice and evaluating at times $t_1$ and $t_2$. This results in 
\begin{align}
    P_{ij}(t_1,t_2) = 
    & \int_{-\infty}^{\infty} dtd\mathbf{x}\int d\vartheta_{1\mathbf{x}}d\vartheta_{2\mathbf{x}} \overline{\nu(\nu-1)}\cdot\overline{F(\mathbf{x})}\chi\left(\vartheta_{1\mathbf{x}}\right)\chi\left(\vartheta_{2\mathbf{x}}\right) \cdot \nonumber \\ &\cdot g_1(\vartheta_{1\mathbf{x}},t\to D_{i},t_1) g_1(\vartheta_{2\mathbf{x}},t\to D_{j},t_2). \label{eq:Joint_prob2}
\end{align}
From here, the Rossi-alpha function, as in Eq.~(\ref{eq:Rossi_a_initial}) is given by:
\begin{equation}
    R_{ij}(\tau) = \rev{\int d\vartheta\bar{\psi}(\vartheta)\sigma_{D,tot}(\vartheta) +} \frac{P_{ij}(t,t+\tau)}{\int d\vartheta\bar{\psi}(\vartheta)\sigma_{D,tot}(\vartheta)}, \label{eq:Rossi_a}
\end{equation}
where we divided by the probability per unit time of getting a count in either of the detectors. As the result  is independent of $t$, one may set $t=0$.

\section{Test Case: Point Model Equations}
\label{sec:test-case}

Our approach holds for any \rev{space- and energy-dependent} stationary subcritical neutron system. Nevertheless, as a simple demonstration of its validity, we turn to the special case of the point model, where there is no spatial dependence. Past works used Langevin equations to calculate the second-order statistics of such models, but the treatment included contributions to fluctuations coming from all physical processes, including the detection itself~\cite{DUBI2018608}. In this section we demonstrate that the second moment statistics of a point model multiplying system (the Feynman-alpha formula) can be reproduced using an effective Langevin equation, accounting only for the branching term, namely fission, as a source to the noise in the system. 

In a point model (in space and in energy), the neutron population is represented by a single time-dependent function $N(t)$. Probabilities for different reactions per unit time are given by the parameters $\lambda_f, \lambda_d, \lambda_a$ for fission, detection and absorption (not including detection), respectively. The total reaction rate is defined as $\lambda = \lambda_f + \lambda_d + \lambda_a$, and a Poisson external source with average emission rate $S$ is present.

The standard point model equation (disregarding fluctuations entirely) describes only the average behavior of the system, and is given as the balance of incoming and outgoing neutrons:
\begin{equation}
 \frac{d}{dt} \overline{N(t)} = \big(-\lambda + \lambda_f \bar{\nu}\big) \overline{N(t)} + S,
\end{equation}
which has the steady-state solution $\overline{N} = S/\alpha$, with $\alpha = \lambda - \lambda_f \bar{\nu}$ the decay constant. The mean number of neutrons detected in a detector in a time interval $[0,T]$ is given by:
\begin{equation}
 \overline{D(T)} = \int_0^T dt \lambda_d \overline{N} = \lambda_d \overline{N} T = \frac{\lambda_d S T}{\alpha}.
\end{equation}
Following our formulation from Section~\ref{comperison}, we define the effective Langevin equation for $\delta n(t)$. Equations~(\ref{eq:delta n_D},\ref{eq:Langevin_final}) reduce in the point model to:
\begin{align}
    \frac{d}{dt} \delta n(t) & = -\alpha \delta n(t) + \sigma \epsilon(t),\\
    \delta n_D(T) &= \int_0^T dt \lambda_d \delta n(t),    
    \label{eq:point_model_langevin}
\end{align}
where the noise coefficient obeys $\sigma^2 = \overline{\nu(\nu - 1)} \lambda_f \overline{N}$ and is completely analogous to the noise coefficient in Eq.~(\ref{eq:source}).  

The formal solution to Eq.~(\ref{eq:point_model_langevin}) is given by,
\begin{equation}
   \delta n(t) = \int_{-\infty}^{t} dt' e^{-\alpha (t-t')} \sigma \epsilon(t') .\label{eq:n_point_model}
\end{equation}
The mean of this contribution vanishes, using $\overline{\epsilon(t)} = 0$. The second moment is given by:
\begin{equation}
    \overline{(\delta n_D)^2(T)}   =  \lambda_d^2 \rev{\int_0^T}\int_0^T dt ds \overline{\delta n(t)\delta n(s)}.
\end{equation}
This double time integral can be evaluated \rev{by substituting Eq.~(\ref{eq:n_point_model}) for $\delta n$ and} using the delta-correlation of the white noise, $\overline{\epsilon(t) \epsilon(t')} = \delta(t-t')$. The result is:
\begin{equation}
    \overline{(\delta n_D)^2(T)} = T \left(\frac{\lambda_d \sigma}{\alpha}\right)^2 \left(1- \frac{1-e^{-\alpha T}}{\alpha T} \right), 
\end{equation}
which leads to the correct result for the Feynman-alpha formula in the point model case, as it appears in \cite{DUBI2018608}, 
\begin{equation}
    Y(T) \equiv \frac{\textrm{Var}\big(D(T)\big)}{\overline{D(T)}} - 1 = \frac{\overline{(\delta n_d)^2(T)}}{\overline{D(T)}} = \frac{\overline{\nu(\nu - 1)} \lambda_f \lambda_d}{\alpha^2} \left(1- \frac{1-e^{-\alpha T}}{\alpha T} \right).\label{eq:feyn_a_pm}
\end{equation}

\section{Discussion and implications} \label{discussion}
\subsection*{Elaborations on the Source Term}
We have established that deviations from Poisson noise of the second moment of the neutron detector reading  can be calculated by assuming it is a Gaussian noise that obeys an Ito-Langevin equation. We suggest that this will help in a full simulation of the noise behavior. To support this claim, we first elaborate on the strength $\sqrt{\overline{\nu(\nu-1)}\cdot\overline{F}}$ of the stochastic source in Eq.~(\ref{eq:Langevin_final}).

One may wonder why the source term is given only by the branching term. After all, the other interactions (i.e. neutron captures and scattering) are also stochastic in their nature and contribute to the reactor noise. Should they not contribute to the noise source? The answer is that these processes contribute only to the Poisson part of the noise. One way to see this is to consider the special case of Poisson detection signal noise at detector $D$ in the time interval $[0,T]$, in the presence of an external Poisson source in a medium with no possibility for fission events. Hence it is a compound Poisson process, and the variance of this noise can be written as equal to its mean,
\begin{equation}
\overline{\delta n_{D}(t_{1})_{\rm Poisson}^{2}}=\overline{N_{D}(S,T)}=\int d\vartheta dtS(\vartheta,t)g_{1}\left(\vartheta,t\rightarrow D_T\right)\label{eq:dn_r1_poisson} .
\end{equation}
To see this, note that in this special case, the Green function in the integrand, $0\leq g_{1}\left(\vartheta,t\rightarrow D_T\right)\leq1$ is exactly the probability that a neutron that is emitted at $\vartheta$ at time $t$ will cause a pulse at detector $D$ at time period \rev{$[0,T]$}. As this is a yes/no question, the distribution is binomial and its variance is $g_{1}(1-g_{1})$. Note that this variance is contributed by all the stochastic process in the medium (scattering and captures) and the stochastic nature of the detection, and takes them into account. In addition to this variance, we must also take into account the contribution of the variance of the external Poisson source itself. We denote the number of neutrons emitted from the source during a time interval $dt$ and into a phase-space volume $d\vartheta$ as $dN(\vartheta,t)$, such that  $\overline{dN(\vartheta,t)}=d\vartheta dtS(\vartheta,t)$. The Poisson source has a variance to mean ratio of one, therefore we also have $\overline{\delta\left(dN(\mathbf{x},\mathbf{v},t)\right)^{2}}=\overline{dN(\mathbf{x},\mathbf{v},t)}$. The overall variance in the number of counts can be calculated using the law of total variance for a compound Poisson random variable, giving the result:\rev{
\begin{eqnarray}
\overline{\delta n_{D}(t_{1})_{\rm Poisson}^{2}} & = & \int \Bigg(\overline{\delta\left(dN(\mathbf{x},\mathbf{v},t)\right)^{2}}g_{1}{}^{2}+\overline{dN(\mathbf{x},\mathbf{v},t)}g_{1}(1-g_{1})\Bigg)\nonumber\\
 & = & \int\Bigg(\overline{dN(\mathbf{x},\mathbf{v},t)}g_{1}{}^{2}+\overline{dN(\mathbf{x},\mathbf{v},t)}g_{1}(1-g_{1})\Bigg)\nonumber\\
 & = & \int \overline{dN(\mathbf{x},\mathbf{v},t)}g_{1} ,
\end{eqnarray}}
which is indeed the same as Eq.~(\ref{eq:dn_r1_poisson}). The variance of the source, with $g_{1}^{2}$ in it, which would have led us to a non-Poisson noise and hence a contribution to the Langevin equation stochastic source, is \textbf{exactly} cancelled by terms from the variance of the one-particle counting process, which takes all the stochastic processes in the medium and detection into account. We can deduce that the only stochastic processes that should be taken into account as contributing to a \textit{non-Poisson} behavior are the branching events.

Another way of interpreting this result is to note that the factor $\overline{\nu(\nu-1)}$ in the noise source in Eq.~(\ref{eq:Langevin_final}) is the numerator of the Diven factor~\cite{diven1956}, and is zero for any other stochastic event in the reactor, such as scattering and captures, in which the number of neutrons emitted is strictly either 0 or 1. As mentioned above, this factor counts the average number of pairs of neutrons emitted from the event, and as such it singles out fission events, which are unique in this aspect. In other words, the numerical factor $\overline{\nu(\nu-1)}$ tells us that the non-Poisson noise is related only to branching processes, which are characterized by a nonzero second factorial moment of the distribution of the emerging neutrons number.

A straightforward but interesting generalization of this result is concerned with a system with more than one species of fissile material, like a (possibly position-dependent) mixture of fissile isotopes, each with a unique $\nu$ distribution. Carrying out the same derivation for this case, we find that for each spatial point $\mathbf{x}$, the noise source is a weighted average of the contributions from all fissile materials 
\begin{equation}
    s(\mathbf{x})=\Big(\overline{\nu(\nu-1)F}\Big)(\mathbf{x}) = \sum_{i}  \overline{F_i(\mathbf{x})} \Big(\overline{\nu(\nu-1)}\Big)_i  , \label{eq:manyF}
\end{equation}
where
\begin{equation}
   \overline{F_i(\mathbf{x})} =   \int d\vartheta_\mathbf{x}  N_i(\mathbf{x}) \sigma_f^i(\mathbf{x})\bar{\psi}(\vartheta_\mathbf{x}) , \label{eq:Fi}
\end{equation}
and where $i$ enumerates the different fissile isotopes, and $N_i$ and $\sigma_f^i$ are the atom number density and the microscopic fission cross-section of the $i$th isotope, respectively. Also, the spectrum is
\begin{equation}
    \overline{\chi}\left(\mathbf{\vartheta}\right) = \frac{\sum_{i}   \overline{F_i(\mathbf{x})}\chi_i(\mathbf{\vartheta})}{\sum_{i}   \overline{F_i(\mathbf{x})}}, \label{eq:manychi}
\end{equation}
where $\chi_i$ is the fission neutron spectrum of the $i$th isotope. It is evident that the noise is composed of separate contributions of the same form from each isotope, with no cross terms such as $\overline{\nu_i\nu_j}$. Again, this result can be traced back to the fact that each detected pair of neutrons that contribute to the non-Poisson part of the noise, has only one \textit{``common ancestor''}, namely one common fission event. 

These explanations also serve as indicators that calculating the deviation from Poisson behavior through our proposed method of a Langevin equation will be much more efficient compared to calculating the whole noise via a brute force Monte-Carlo method. In the latter, one would have to spend a lot of calculation time and resources to take into account the possibility of every single interaction of the neutrons in the reactor in order to gather enough statistics to accurately reproduce the mostly-Poisson behavior created by most reactions occurring. This is instead of calculating only the smaller addition to the deviation from Poisson statistics that we are really after. In our approach, we calculate the deviation directly, requiring significantly fewer computational resources. This would be especially true for cases of low detection efficiency, which leads to small values of $Y(D,T)$ in Eq.~(\Ref{eq:Fenymann Y}). 

\subsection*{Numerical Implications}
We now present a scheme to efficiently calculate the non-Poisson part of the noise, $d_{ij}$, according to our method. To do that, one would divide the fuel region (where the fission rate $\overline{F(\mathbf{x})}$, and hence our effective source, are non-zero) into $M$ small segments of nearly constant $\overline{F}$. \rev{This method introduces a systematic error if the minimal size of a segment is larger than the length scale of change in the neutron flux or fission density. If the segmentation becomes finer than this scale (which is of the order of a few centimeters - the diffusion length of the system), the systematic error can be made arbitrarily small (the resulting expressions converge to the exact result as the segmentation becomes finer). Note, however, that increasing the number of segments while keeping the total number of histories constant would not increase the statistical error significantly, and it might even help to reduce it, when using methods of variance reduction \cite{NonAnalogousMC2014}}. The division can be according to fuel elements, parts of fuel elements, or to a particular grouping of fuel elements. For each segment $m\in\{1,..,M\}$, one would then assign the amplitude of the stochastic source, $s_m\equiv\int_{\mathbf{x}\in m}{d\mathbf{x}\big(\overline{\nu(\nu-1)F}\big)(\mathbf{x})}$, by solving once a standard time-independent source problem to find $\bar{\psi}(\mathbf{x})$, then evaluate the integrand using Eq.~(\ref{eq:manyF}). 

In principle, one can then try to use the Langevin method and solve for many realizations of the noise function $\epsilon(\mathbf{x},t)$. \rev{This approach will be useful in solving for the noise in discrete models of neutron transport, and will increase the numerical efficiency as a result of the fewer noise terms in this formulation, compared to the other approaches mentioned in Sec.~\ref{sec:intro}. In three-dimensional, heterogeneous problems, o}ne may also expedite the calculations by solving, using a standard time-dependent transport code (either Monte Carlo code or a deterministic one), for $g_{1}(\vartheta,t\rightarrow D_{i},t')$ for multiple fission points as the starting point $\vartheta$. Defining the Green functions on each segment as
\begin{equation}
    \tilde{g}_{1}(m,t\rightarrow D_{i},t')\equiv \int_{\mathbf{x}\in m}d\vartheta \chi_m\left(\mathbf{\vartheta}\right)g_{1}(\vartheta,t\rightarrow D_{i},t') ,
\end{equation}
and 
\begin{equation}
    \tilde{g}_{1}(m,t\rightarrow D_{i,T})\equiv \int_{0}^{T}dt'\tilde{g}_{1}(m,t\rightarrow D_{i},t') ,
\end{equation}
and then numerically integrating Eq.~(\ref{eq:d_ij_g1_g1}) using the approximation of homogeneous source $s_m$ at each segment $m$, leads to the sum
\begin{equation}
   d_{ij}\sim  \sum_{m=1..M}{ s_m\rev{\int_{-\infty}^{\infty}} dt\tilde{g}_{1}(m,t\rightarrow D_{i,T})\tilde{g}_{1}(m,t\rightarrow D_{j,T})}\label{eq:dijm}.
\end{equation}
Numerically integrating Eq.~(\ref{eq:Joint_prob2}) leads to
\begin{equation}
   P_{ij}(t_1,t_2)\sim  \sum_{m=1..M}{ s_m\rev{\int_0^\infty} dt\tilde{g}_{1}(m,t\rightarrow D_{i},t_1)\tilde{g}_{1}(m,t\rightarrow D_{j},t_2)}\label{eq:Pijm}.
\end{equation}
\rev{To treat the temporal integrations in the above expressions, one may calculate $\tilde{g}_1$ on a discrete set of time differences $t'-t$ and then use any standard numerical integration scheme.}
Equations~(\ref{eq:dijm}) and (\ref{eq:Pijm}) provide an efficient route to calculate the non-Poisson noise correlations $d_{ij}$ and the probability \rev{density} function of \rev{correlated neutron pair} detection $P_{ij}$, which can be used  for calculating Feynman-alpha and Rossi-alpha functions, respectively.

\rev{We can estimate} the improvement in efficiency by using these schemes, for instance, in means of number of neutron histories needed in a Monte-Carlo simulation to calculate $d_{ij}$ with a certain statistical error. \rev{We compare our method to analogue Monte Carlo schemes, which are commonplace in simulating neutron noise (e.g. the analogue capture module in MCNP6 \cite{werner2017mcnp}). Importantly, this estimate is exact in the case of the point model, namely when trying to simulate a subcritical system with space- and energy-independent flux.} We focus on the case where the detection efficiency, in neutron detection per fission, denoted $\mu$, is very small (i.e. detection is rare), and also where the distance from criticality $1-k$ ($k$ is the effective multiplication factor) is substantial enough that the fission chains are not too long. These conditions ensure that the non-Poisson part of the noise is dominated by two detected neutrons from the same fission chain at most, and  we can neglect the contributions of three or more pairs of detected neutrons from the same chain. This is a common situation for sub-critical assemblies, safeguards multiplicity measurements and nuclear reactor experiment in start-up, and as we shall now see, it is the case where our method has the most substantial advantage. For simplicity, we will focus on the statistical error in predicting the long time-period limit of the noise, although the relative improvement in efficiency is expected to hold for general~$T$. 

We first estimate the number of histories needed for predicting the non-Poisson part of the noise in a standard Monte-Carlo neutron simulation, \rev{and then compare this to the number of histories required in our scheme. We will show that these estimations are \textbf{exact} in the point model, and argue that the increased efficiency should hold, to within an order of magnitude, in a general 3-dimensional heterogeneous transport calculation. Supposing we start a point model Monte-Carlo simulation with $N_s$ external source neutrons. These neutron will produce on average $\bar{N_h}=N_s/(1-k)$ Neutrons (including source neutrons), each will have its own history simulated. Thus $\bar{N_h}$ will be the average total number of histories in such a simulation, and $\bar{D}=\bar{N_h}\frac{\lambda_d}{\lambda}$ of them will end up detected. If we now use the above made assumption that each source neutron usually leads to the detection of at most one correlated pair, then the detected pairs are not correlated, and if we denote $\bar{D}_{2}$ the average number of detected pairs in the simulation, we can assume that the statistical error of the number of pairs equal to its average $Er^2=\bar{D}_{2}$.}

\rev{We calculate $\bar{D}_{2}$ following Eq. \ref{eq:feyn_a_pm} (and taking a very long measurement interval), and the known relation between the detected pairs and the non Gaussian part of the variance, $Y_\infty=2\frac{\bar{D}_{2}}{\bar D}$ \cite{CROFT2012152}. The average total number of neutron pairs counted in this simulation is given in the point model by
\begin{equation}
    \bar{D}_{2} = \overline{\nu(\nu-1)}\frac{N_s\lambda_f\lambda_d^2}{2\alpha^2\lambda}=\overline{\nu(\nu-1)}\frac{N_s\lambda_f^3\mu^2}{2\alpha^2\lambda}, \label{eq:pair_number_pm}
\end{equation}
where we identify the detection efficiency, in units of counts per fission, as $\mu=\frac{\lambda_d}{\lambda_f}$. Next we use the point model relation $k=\frac{\bar\nu\lambda_f}{\lambda}$ to rewrite this equation in terms of the multiplication factor. The term
\begin{equation}
    \frac{\lambda_f}{\alpha} =\frac{1}{\bar\nu}\frac{k}{1-k},
\end{equation}
is the average total number of fissions occurring per source neutron. Substituting  in \ref{eq:pair_number_pm}, and using the above mentioned relation between $N_s$ and $\bar{N_h}$ we get
\begin{equation}
    Er^2=\bar{D}_{2} = \frac{1}{2} \overline{\nu(\nu-1)} \cdot \bar{N_h} (1-k) \frac{k}{\bar{\nu}} \cdot \bigg(\frac{1}{\bar\nu}\frac{\mu k}{1-k}\bigg)^2.
\end{equation}
This expression can now be understood intuitively, in the spirit of our work - the first term in the product is the average number of neutron pairs emitted per fission, the second is the average total number of fissions in the simulation, and the third is the square of the point-model "Green function" - the number of detected neutrons per emitted neutron (cf. Eq.~\ref{eq:dijm}).} 

Hence, for a required relative error $\mathrm{Er}$ we would need $\frac{1}{\mathrm{Er}^2}$ detected pairs, which leads to a number of required neutron histories of 
\rev{\begin{equation}
N_h = \frac{2(1-k)\overline{\nu}^3}{k^3\mu^2\overline{\nu(\nu-1)}\mathrm{Er}^2}.\label{eq:their_Nh}    
\end{equation}}
For typical values of the parameters in the case of research reactors \cite{CROCUS2021, Jordanian2017}, we take $\mu=10^{-4}-10^{-5}$, $\overline{\nu}=2.43$, \rev{$\overline{\nu(\nu-1)}=4.63$}, $\mathrm{Er}=0.01$, and  $k=0.95$, we find $N_h\sim3\times 10^{11}-3\times 10^{13}$ histories, which are exceptionally costly and close to the limit of current computation power \cite{mickus2018optimal, hoogenboom2011monte}.

In our method, for an accurate estimation of Eq.~(\ref{eq:dijm}) one needs to calculate accurately the product of two Green functions $\tilde{g}_{1}(m,t\rightarrow D_{i,T})\tilde{g}_{1}(m,t\rightarrow D_{j,T})$ (say, for $i\neq j$). On average (i.e. ignoring position-dependent deviations), calculating the Green function for a long time period $T$ is equivalent to calculating the efficiency per emitted neutron. For an average value $\mu/\bar{\nu}$, and given that detected neutrons in the simulation for the above case are mostly uncorrelated, the relative statistical error of the Green function would be about $\frac{1}{\sqrt{\mu/\overline{\nu}N_h}}$. Therefore, for a required total relative statistical uncertainty $\mathrm{Er}$ of the product of the two Green functions, one finds the needed number of Monte-Carlo histories of emitted neutrons from fission events to be 
\begin{equation}
N_h\sim \frac{2\overline{\nu}}{\mu\mathrm{Er}^2}\label{eq:our_Nh}  
\end{equation}
For the same values of the parameters given above, we now have $N_h\sim5\times10^8-5\times10^9$ histories, which is achievable in present computational resources. Note the difference by three orders of magnitude between the results of Eq.~(\ref{eq:their_Nh}) and  Eq.~(\ref{eq:our_Nh}), which is mainly due to the difference in the dependence on detection efficiency $\mu/\overline{\nu}$ -- linear dependence in our case vs.\ quadratic dependence in the standard calculation. 

In addition, studying individual terms of the sum in Eq.~(\ref{eq:dijm}), our method gives useful information about the position-dependent contribution of each of the $M$ segments to the non-Poisson noise correlations. In particular, this analysis could be useful for the design and optimization of safeguards assay tools: calculating this spatial distribution of contributors to the noise will inform on the system's effective field of view, assaying the fissile material more efficiently. Note that this contribution distribution may have dependence on the time period $T$, which may contain more information. 

Going beyond these estimates, \rev{we first note that for a heterogeneous system there may be variations in the efficiencies stated above, but for the above sub-critical systems, whose observed deviations from the point-model are not too significant, the rather large gain in efficiency is expected to hold. Indeed,} the actual number of required histories may change substantially depending on the specifics of the system and of the fidelity required; fully predicting the time-dependent features requires more histories than the prediction above. \rev{These temporal features are needed, for example,} for characterizing higher decay modes of the neutron population, which are dominant on shorter timescales. However\rev{, note that calculating these temporal effects usually involves filtering out histories, which effectively means reducing the detection efficiency $\mu$, which further increases the relative computational gain in our method;} we believe that in all of these cases the method presented above has a clear advantage in computation time over standard noise calculations.  


\bibliographystyle{unsrt}
\bibliography{refs}

\end{document}